# An atomic resolution, single-spin magnetic resonance detection concept based on tunneling force microscopy

A. Payne, K. Ambal, C. Boehme*, C. C. Williams**

Department of Physics and Astronomy, University of Utah, 115S 1400E, Salt Lake City, UT 84092, USA

A comprehensive study of a force detected single-spin magnetic resonance measurement concept with atomic spatial resolution is presented. The method is based upon electrostatic force detection of spin-selection rule controlled single-electron tunneling between two electrically isolated paramagnetic states. Single spin magnetic resonance detection is possible by measuring the force detected tunneling charge noise on and off spin resonance. Simulation results of this charge noise, based upon physical models of the tunneling and spin physics, are directly compared to measured AFM system noise. The results show that the approach could provide single spin measurement of electrically isolated qubit states with atomic spatial resolution at room temperature.

# I. INTRODUCTION

Paramagnetic point defects in semiconductors are among the most coherent qubits found in nature [1, 2], yet their utilization requires reliable single-spin readout techniques which allow to access the individual defect states with atomic resolution. Spatially well-defined single spin-readout utilizing spin-selection rules has been demonstrated in the past on electronic transitions between double charge quantum dots [3-5]. Applying a similar approach to the spin measurement of individual paramagnetic point defects has been proposed [6, 7], yet after more than two decades since the first single-spin detection experiments [8], the spatial resolution of various electrical [9-11], optical [12] and even scanning probe based single spin detection techniques [13-15] is one to two orders of magnitude above the localization of the paramagnetic states [8, 9, 11, 13, 14]. This limitation makes the application of these spin measurement techniques for a selective readout of adjacent paramagnetic states difficult – or, as recently demonstrated [15], they are based on scanning tunneling microscopy, requiring currents and conducting substrates that could limit spin coherence times of qubits when the spin readout is used for quantum information applications.

In recent years, individual electronic tunneling events have been observed by single electron tunneling force microscopy which is based on the detection of electrostatic forces caused by single electron tunneling between electronic point defects and a conducting AFM cantilever probe [16]. Its capabilities for single electron tunneling spectroscopy, imaging and quantum state depth measurement on the atomic length scale have been well demonstrated [17-23]. Since this method relies on electrostatic force detection of individual tunneling events, it can operate on surfaces that are completely non-conductive.

## II. A SPIN-SELECTION RULE, FORCE MICROSCOPY BASED SINGLE-SPIN READOUT CONCEPT

For single-spin readout with atomic spatial resolution, we propose here a combination of single electron tunneling force microscopy with magnetic resonance, conceptually illustrated in Figure 1(a). It consists of a scanning probe AFM tip made out of a weakly spin-orbital coupled material (e.g. silicon or silicon dioxide) whose tip contains a single paramagnetic electron state, the probe spin, as well as mutually perpendicular $B_0$ and $B_1$ field coils for DC and RF magnetic field generation needed to establish the magnetic resonance conditions for the electron spin resonance. Figure 1(b) illustrates three spin- and charge occupation scenarios which can occur when the probe spin is brought, both energetically and spatially, into tunneling range of a test spin. When the spin-pair state of the two centers have high permutation symmetry (high triplet content), the Pauli blockade caused by the weak spin-orbit coupling quenches the electron tunneling probability between the tip and the surface. In contrast, when the two centers form a pair with high permutation anti-symmetry (high singlet content), the tunneling probability is finite and randomly occurring electron tunneling events can be expected between the two centers producing a surface charge random telegraph signal (RTS) and a corresponding frequency shift on the oscillating AFM probe.

In order to detect individual spin states, we propose here to observe the change in the electrostatically induced cantilever frequency noise generated by the random tunneling transitions between the test spin and the probe spin when at least one of the two spin states are brought into magnetic resonance.

## III. PROPERTIES OF SUITABLE PARAMAGNETIC ELECTRONIC STATES

For the paramagnetic probe states involved in the single-spin detection concept, we suggest using a silicon dangling bond state in the amorphous $SiO_2$ network, (so called E'$_\gamma$ center) as it exhibits many properties needed for the single spin detection concept described in this study (24, 26): 1) the $E'_\gamma$ center is highly localized (few Angstroms) [27], providing a localization range needed for atomic-scale spatial resolution. 2) $SiO_2$ is a good dielectric due to its large band gap. The $E'_\gamma$ center is energetically rather deep in the gap [27], so that an electron injected into the defect can remain there for a long time. 3) $SiO_2$ films can be easily grown on standard silicon AFM tips by thermal oxidation. 4) $E'_\gamma$ centers have long spin-lattice relation times ($T_1$) of $\approx$200 µs at room temperature [28, 29]. There are many variations of $E'_\gamma$ centers found in amorphous $SiO_2$ due to the large spread of bond angles and bond lengths found in the amorphous network [30-34]. While the microscopic theory for the $E'_\gamma$ center is not without controversy [35], the common feature among all variations is an unpaired electron on a silicon atom back-bonded to three oxygen atoms, i.e. a silicon dangling bond. The electronic structure of the dangling bonds associated with oxygen vacancies have been calculated by several groups [27, 36]. In this work, the $E'_\gamma$ center is taken as a prototypical paramagnetic defect for the simulations of the proposed single electron spin detection method. We assume in this simulation that tunneling occurs between two $E'_\gamma$ centers, one in an ized silicon probe tip and the other at an oxidized sample surface.

## IV. SIMULATION OF FORCE DETECTED SINGLE-SPIN DETECTION

Random electron tunneling between two weakly spin-orbit coupled electron states, one located on a cantilever tip and the other at a sample surface, produces a random telegraph charge signal governed by two different, random, spontaneous processes: electron tunneling and spin flips. The spin flips may be driven by either intrinsic longitudinal spin relaxation (so called $T_1$ processes) or magnetic resonance (determined by the resonant driving field $B_1$). The expectation value $T_t$ of the tunneling time (the average time the electron stays in one state before it tunnels to the other state) depends on the height of the energy barrier and the width of the gap between the states. Both tunneling and spin relaxation are independent stochastic processes and obey Poisonian statistics. If the field strength $B_1$ of the applied rf radiation is large and its frequency meets the magnetic resonance condition $\Upsilon B_0 = hf$, with $\Upsilon$ being the gyromagnetic ratio and $h$ the Planck constant, the spin flip rate may be high, with an average flip time of $T_{flip} = 1 / (\Upsilon B_1)$ much less than the intrinsic longitudinal spin relaxation time $T_1$. Under this condition, the relatively slow "blinking," or pausing of the tunneling (caused by Pauli exclusion) on the time scale of $T_1$ is eliminated, reducing the low frequency component of the charge fluctuation at the surface (charge noise).  It is important to note that the on-resonance spin flipping are still stochastic and follow the probability distribution discussed earlier, but the average flip time is small compared to $T_1$. The average tunneling time ($T_t$) does not change when magnetic resonance is achieved since it is determined only by the energy barrier.

The simulations are started with the two paramagnetic states separated (one electron in the probe state of the tip and the other in the test state of the sample) in an antiparallel spin configuration.  Using the probability distributions for each process, a random number generator then produces a random tunneling time, as well as a random spin flip time at each time step. If a spin flip occurs, the spin-pair permutation symmetry changes from either parallel to mixed or vice versa. If the spin pair is in a parallel configuration, tunneling is blocked until another spin flip occurs.  If the spin pair is in a mixed spin configuration, electron tunneling is allowed, creating a doubly charged singlet state, in which no spin flipping can occur.  Either of the two electrons can then tunnel back to the tip state producing a separated charge state with mixed spin permutation symmetry. The simulation thus creates a transient series of tunneling events to and from the sample state, represented by a 0 for the separated electrons and 1 for the doubly charged state in the tip. Simultaneously, they produce records of the relative spin orientation of the two electrons for each time step.

The simulation produces step like transitions in the time domain, corresponding to an infinite bandwidth.  In real experiments, the force detection of the tunneling charge occurs with a finite bandwidth. To include this effect, an adjustable, first-order band pass filter is implemented in the simulation code (Butterworth filter with 3dB roll-off) to take into account the finite experimental bandwidth effect.

Single spin detection will require tunneling rates $T_t^{-1}$ much higher than the spin-lattice relaxation rates $T_1^{-1}$ of either the test or the probe spin. As long as this condition is met and $T_1^{-1}$ is smaller than the magnetic resonantly induced spin flip rate coefficient ($\sim \gamma B_1$ when $B_1$ is sufficiently large), the qualitative results of this simulation will be equally applicable to other types of probe and test spins. As tunneling times depend sensitively on the distance between the probe and test spin, which can be well controlled on an Å-scale with state of the art scanning probe setups, we can assume that such a high tunneling rate for a non-Pauli blocked spin state is established.

For the simulations, tunneling and spin-flip transient events at discrete time steps of 100ns were generated. The chosen times steps are small compared to all other important physical processes (the tunneling time expectation value $T_t$, the intrinsic spin relaxation time $T_1$ = 200µs for each of the spin pair partners as well as the resonantly driven average spin flip rate $T_f \approx$ 10µs in presence of magnetic resonance).

## V. INDEPENDENCE OF AVERAGE CHARGE IN PROBE AND TEST STATES ON MAGNETIC RESONANCE CONDITIONS.

It is important to establish whether the average charge in one of the states depends on the magnetic resonance condition, since this would provide a simple, direct method to detect the spin-resonance condition for a single spin. This is particularly true, since the AFM is capable of detecting changes in surface charge with sub-electron sensitivity [16]. However, a simple rate picture shows that the average charge in the test or probe states does not depend on whether the magnetic resonance condition is reached as described below.

Figure 1(b) illustrates three of the five charge/spin configurations that two singly occupied, weakly spin-orbit coupled paramagnetic states can assume when their energetic and spatial alignment allows for tunneling. The sketch on the left represents the two charge separated states in which the spin-configuration is in either one of two pure triplet states, the $|\uparrow\uparrow\rangle = |T_+\rangle$ or the $|\downarrow\downarrow\rangle = |T_-\rangle$ state. When spin-flip occurs due to intrinsic relaxation (a $T_1$ process) or due to a magnetically resonant excitation, the charge separated triplet state will change into one of two charge separated product spin-states with singlet content, either the $|\uparrow\downarrow\rangle$ state or the $|\downarrow\uparrow\rangle$ state, illustrated by the center sketch of Figure 1(b). A transition of a charge separated triplet state (left) into the doubly occupied charged state (right), which can only exist in singlet configuration (illustrated on the right) is not allowed due to spin-conservation. However, a transition of the mixed permutation-symmetry state (center) into the doubly occupied state is allowed. Note that at small tip-sample separations, the charge separated electron states will be weakly spin-spin coupled due to the exchange- and dipolar-interaction between the two spins. The spin configuration of the pair is therefore always one of the four product states of a 2 spin s=1/2 system. Thus, Figure 1(b) represents a rate system consisting of charge separated states with pure triplet (two states represented by the sketch on the left) and mixed spin-permutation symmetry (two states represented by the center sketch), and the doubly charged state with singlet spin-configuration (represented by the sketch on the right). Note that establishing magnetic resonance for either one of the four charge separated spin states will increase the average spin flip rate from its intrinsic value determined by the spin relaxation rate $T_1^{-1}$ to the magnetic resonance driven average rate controlled by the driving field strength $B_1$.

Spin relaxation transitions only change spin states but not the charge state. The average steady state occupation probabilities for all five states of the given rate system are therefore independent of the spin flip rate. For the four charge-separated states these probabilities are 1/6, for the doubly charged state it is 1/3. Measurements of the average charge in the test state and probe state reveals solely the average occupation probability of the doubly occupied singlet state and the sum of the occupation probabilities of the four product spin states, which are $e$/3 and 2$e$/3, respectively. These values are independent of whether magnetic resonance is present or not and thus, average charge measurements are not suitable for single-spin detection.

## VI.     SINGLE SPIN DETECTION OBSERVABLE

Figure 2(a) displays the plot of the simulated charge in the spin states ("0" and "1" indicating the separated/non-separated charge cases, respectively) as a function of time during the first 2ms of a 100 ms simulation for the two cases of the presence (blue) and absence (red) of magnetic resonance of at least one of the spin pair partners. For the off-spin resonant case, the RTS includes blinking (time periods of finite and zero tunneling rates). The finite tunneling rate occurs as an electron tunnels back and forth between a doubly occupied singlet state [Figure 1(b), right] and the separated product state [Figure 1(b), center]. When the separated product state undergoes a $T_1$ relaxation (a spin flip) into the separated triplet state [Figure 1(b), left], Pauli blockade is present and no electron tunneling occurs. In contrast, in presence of a continuously applied strong magnetic-resonant driving field $B_1$, tunneling between the separated singlet and triplet state can occur frequently. Consequently, the blinking in the tunneling dynamics vanishes.

The simulations also reveal that the change of the spin-dependent tunneling dynamics between on- and off- resonance does not affect the average charge (2/3 in both cases) in the probe or sample spin state, as indicated by the dashed lines in Figure 2(a) and consistent with the arguments in section IV above. In contrast to the average charge in the probe or test states, the dynamics of the random tunneling transitions (charge noise) between the probe and test states does provide a measurable signal for detection of the magnetic resonance condition and thus, the detection of a single spin.

As can be seen in the full 100 ms simulation results, the RMS value and noise power spectral density of the tunneling charge variation in either state is affected, as plotted in Figure 2(b) and 2(c) by the off- (red) and on- (blue) magnetic resonance condition. The two spectra show that in the absence of magnetic resonance when $T_1$ processes influence the tunneling dynamics, intensive low-frequency noise contributions appear compared to the on-magnetic resonance case. Detection of magnetic resonance of either the probe or the test spin or both (note that even flipping both pair partners at the same time changes the spin-permutation symmetry of the pair [37]) can be determined by measurement of the noise power (RMS) of the RTS signal within an appropriate detection bandwidth.

In the next section, the simulation results show the tunneling charge noise (RMS) has been calculated as a function of $B_1$ strength and applied RF frequency.

### VII. RMS CHARGE NOISE DEPENDENCE ON RF FREQUENCY AND MAGNETIC FIELD STRENGTH

In Figure 3, the RMS tunneling charge noise is plotted as a function of rf frequency at rf magnetic field strengths ($B_1$) varying from 0.1 µT to 100 µT, for a static magnetic field ($B_0$) of 5 mT. To perform these simulations, the rf magnetic field driven spin flip rates are calculated as a function of frequency and $B_1$ using Rabi's formula [38]. The average spin flip times are then used in the simulations to predict the RMS charge noise as a function of rf frequency. At low amplitudes of $B_1$, the resonance peak is indistinguishable from the off resonance RMS noise level. As $B_1$ increases, the magnetic resonance signature increases its signal to noise ratio and is also power broadened. The power broadening increases the width of the resonance peak and therefore makes it easier to find.

### VIII. RELATION BETWEEN SIMULATED CHARGE NOISE AND CANTILEVER FREQUENCY SHIFT.

In order to determine whether the predicted charge noise variation due to spin resonance is detectable, a comparison with the charge detection sensitivity of an actual AFM system is required. A theoretical model, illustrated in Figure 4, is used to calculate the change in frequency shift of an oscillating AFM cantilever under a given set of experimental parameters caused by a single electron tunneling event between states in the tip and sample. This calculation is then used to properly scale the simulation results (with output 0 or 1) to an actual frequency shift of the AFM cantilever for those experimental parameters, to be compared with actual measurement noise. To determine this scaling factor, the change in the electrostatic force gradient on the tip oxide produced by an electron tunneling from a singly occupied defect state in the tip oxide (probe state) to a doubly occupied defect state in the surface (test state) of the sample oxide is calculated based upon the coulomb interaction between point charges. It is assumed that the depth of both states is 0.2 nm, which is small compared to the oxide thicknesses (tip oxide: 10 nm and sample oxide: 15 nm), so that image charge effects in the conducting plane behind the oxide can be neglected. The two defect states are schematically sketched in Figure 4 along with the relevant electrostatic parameters.

The electrostatic force gradient is calculated for the two occupancy cases (charges separated and together in the sample state) as a function of vacuum gap, oxide thickness, depth of each state and external voltage bias, assuming a dielectric constant of 3.9 for the silicon dioxide films. This electrostatic force gradient is then converted to an AFM cantilever frequency shift [39] using experimental AFM parameters: spring constant $k$ = 40 N/m, resonance frequency $f$ = 311745 Hz, quality factor $Q$ = 6441, oscillation amplitude *A* = 10 nm and an applied voltage of *V* = 10 V. Using these values, the magnitude of the frequency shift caused by a single electron tunneling event (scaling factor) is calculated to be between 11.4 Hz and 13.0 Hz for tip-sample gaps ranging between 0.62 and 0.052 nm. These scaling factors (at different tip-sample gaps) are employed in the scaling of the simulated charge noise to an AFM frequency shift noise.

## IX. ENERGY ALIGNMENT OF PROBE AND TEST STATE.

In order for a localized electron in a paramagnetic point defect at the tip of a cantilever to tunnel elastically to a defect state in the sample surface, an energy alignment condition must be met, i.e. the energy of the electron in the singly occupied tip state must be equal to the energy of the sample state when doubly occupied. This implies that the energy of the singly occupied tip state must be higher than the singly occupied sample state, by an energy Δ, equal to the energy difference between the singly and doubly occupied state. When this energy condition is met, an electron can randomly tunnel back and forth between the two states (at finite temperature) with a tunneling rate that is governed by the tunneling barrier height between the two states and the distance between them.

Figure 5 illustrates this energy requirement for the two paramagnetic states. The solid and dashed horizontal lines represent the energy of singly and doubly occupied states, respectively, while the Gaussian curves represent the probability distribution of the energies of these singly and doubly occupied states for a given representative material system. Note that the singly occupied state on the left (solid line) must be energetically aligned with the doubly occupied state on the right (dashed line). The width of the distribution of energies in the tip and sample determines the likelihood of finding two states that meet this energy condition without an externally applied electric field. Tuttle and collaborators have shown that the energy difference between the singly and doubly occupied dangling bond E'$_\gamma$ defect is approximately 1eV [40].

Since two randomly chosen states in the sample and tip oxides may not have the appropriate energies for elastic tunneling between them (the energy of the singly occupied tip state not equal to the doubly occupied sample state), an external voltage bias is needed to bring these two states into energy alignment. With 10V applied between the back contacts of the oxide films, the energy of one state can be shifted relative to the other. Only part of the applied voltage is dropped between the two defect states. Under the experimental conditions described above, a relative energy shift between 0.24 and 1.0 eV is provided by an applied voltage of 10 volts for tip-sample gaps between 0.052 and 0.62 nm respectively. The electric field in the oxide films with this applied voltage is small compared to the breakdown field of silicon dioxide [41].

## X. ROOM TEMPERATURE AFM SYSTEM NOISE.

The ability to detect a single spin depends upon whether the experimental system noise on the AFM frequency shift is smaller than the frequency shift noise for the on and off magnetic resonance cases as calculated. The frequency shift noise of a room temperature Omicron UHV AFM Multiprobe S has been carefully measured as a function of tip sample gap, bias voltage and cantilever oscillation amplitude for comparison with properly scaled simulation data.

For the AFM system noise measurements, a 15 nm oxide film was thermally grown on a standard silicon AFM probe tip. The cantilever was then back coated with aluminum in order to increase its reflectivity. The oxide thickness on the tip was estimated by simultaneously growing an oxide on a planar silicon wafer and measuring it with an ellipsometer. An oxide film was also grown on a silicon sample and measured with an ellipsometer to have a thickness of 10nm. The oxide films were cleaned in the UHV AFM chamber using a heat treatment of 600C for 1 hour for the sample and 250C for 12 hours for the probe tip.

The absolute tip sample gap is a critical parameter in calculating the scaling factor introduced above which is required to compare the simulation results with the measured AFM system noise. The damping and amplitude channels are used to determine the absolute tip-sample gap. Figure 6 shows a typical *df*(z) curve, along with the corresponding dissipation and oscillation amplitude data as a function of the gap (*z*). As the tip approaches the sample surface, the dissipation signal remains constant even in the presence of changing frequency shift (*df*), as expected. At approximately 0.3 nm from the minimum of *df*(z), the dissipation signal sharply increases. This sharp increase in the dissipation signal is attributed to the apex of the probe making significant repulsive contact with the sample surface, causing the dissipation signal to increase sharply [42]. To perform the AFM system noise measurements, the power spectral density of the noise of the frequency shift *df* was measured as a function of the tip-sample gap. In these measurements, the *df*(z) curve in Figure 6 was used for the determination of the tip sample gap, and the contact point (zero gap) was determined by the z value at which the sharp rise in the dissipation signal occurred.

Calculations have been performed [43] that show that the average tunneling rate between the probe and test state is much faster than the inverse spin relaxation time (1/200μs), for state depths of 0.2nm and tip-sample gaps ranging between 0.05nm and 0.62nm, corresponding to the values from the AFM noise measurements.

## XI. COMPARISON OF MEASURED AND SIMULATED NOISE.

The results of these simulations are displayed in Figure 7(a) displays the results of these simulations for two rf frequencies corresponding to the off- (red) and on- (blue) magnetic resonance cases. In this plot, the simulated charge noise (RMS) was converted to frequency shift (RMS) using the electrostatic calculation described in a previous section. The difference between the RMS frequency shift noise for these two cases is significant. While these data where obtained for realistic simulation parameters, they do not account for the presence of additional system noise found in real AFMs, which must be appropriately taken into account to determine whether the approach is viable for single spin detection. The black data points, taken at various tip-sample gaps, represent the experimentally measured room temperature AFM frequency shift system noise as a function of detection bandwidth in the modified [44] commercial scanning probe microscope previously mentioned in Section IX. The measurements were performed with an applied voltage of 10V and consequently the obtainable energy shift between two states are calculated and shown in the table of Figure 7(a). For larger detection bandwidths (>1000 Hz), the AFM system noise exceeds the on-magnetic resonance frequency shift noise and even approaches the off-magnetic resonance RTS noise power. Similarly, as seen from Figure 7(a), at very small bandwidth, the difference between on- and off- resonance becomes small, leading to a detectability loss of the magnetic resonance signal. However, for the given simulation parameters and the measured noise data, there is a bandwidth range between 10Hz and 1kHz in which the system noise is significantly lower than the off-resonant RTS noise. Hence, for the given spin relaxation- and tunneling-parameters, the given scanning probe setup and bandwidths, force detected single spin magnetic resonance detection becomes possible at room temperature.

In a single spin detection experiment, the frequency of the applied rf magnetic field is swept through magnetic resonance. Figure 7(b) shows how the RMS value of the frequency shift noise power as a function the frequency of an applied rf field can reveal magnetic resonance of a single spin, in the presence of real AFM system noise. In these calculations, the AFM system noise power RMS has been appropriately added to the charge tunneling frequency shift noise power (assuming it is uncorrelated, i.e. sum of the squares). The error bars in these plots represent the standard deviation of RMS fluctuations obtained from multiple simulations of 1000ms length and calculated variations of measured experimental noise, assuming Gaussian statistics. The standard deviation of the RMS AFM system noise was obtained by simulating a Gaussian noise power spectrum which was matched to the measured RMS value of the AFM noise measurements. The three data sets show the detectability of magnetic resonance for several detection bandwidths.

Finally, we have studied the effect of frequency-sweep rates (which determine the integration time per frequency data point) on the measured single spin detection signal to noise ratio. Figures 8(a) and 8(b) show simulations of frequency sweep measurements for a 100 and 10 ms integration time per frequency step respectively. These curves can be directly compared with the results shown in Figure 7, which assumes a 1 second measurement time per frequency step and holds all other parameters the same. Figure 8(a) shows that the spin resonance spectrum can be detected with significant signal to noise ratio (S/N) when a 1000Hz detection bandwidth is used and the rf frequency is swept at an acquisition time of 100ms per frequency step. For smaller detection bandwidths and shorter acquisition times, the S/N decreases. Figure 8(b) was simulated with an assumed acquisition time of 10 ms per frequency step. The data sets in Figure 8 show that single spin magnetic resonance detection can be achieved at room temperature for frequency scan rates as fast as 10 ms per frequency step.

## XII. CONCLUSIONS AND SUMMARY

From the comparison of the physically based simulations presented here and measured AFM system noise, it is concluded that a combination of force detected tunneling and magnetic resonance, spin-selection rule based single-spin detection is possible with atomic spatial resolution on electrically isolated paramagnetic states. An experimental demonstration of this concept includes several technical challenges including light-free scanning probe detection to prevent optical excitation of paramagnetic states (possibly by using quartz tuning forks [45]); appropriate management of static and oscillating magnetic fields in a scanning probe setup as well as the development of silicon scanning probes with an accessible, highly localized $E'_\gamma$ state with long spin-relaxation times in an thin silicon dioxide layer near the tip apex [46].

**Acknowledgments:**

This work was supported by the Army Research Office, grant #W911NF-10-1-0315. The contributions by K.A. were supported by the National Science Foundation through MRI project #0959328 and the Utah MRSEC center, grant #1121252.

\* boehme@physics.utah.edu
\*\* clayton@physics.utah.edu

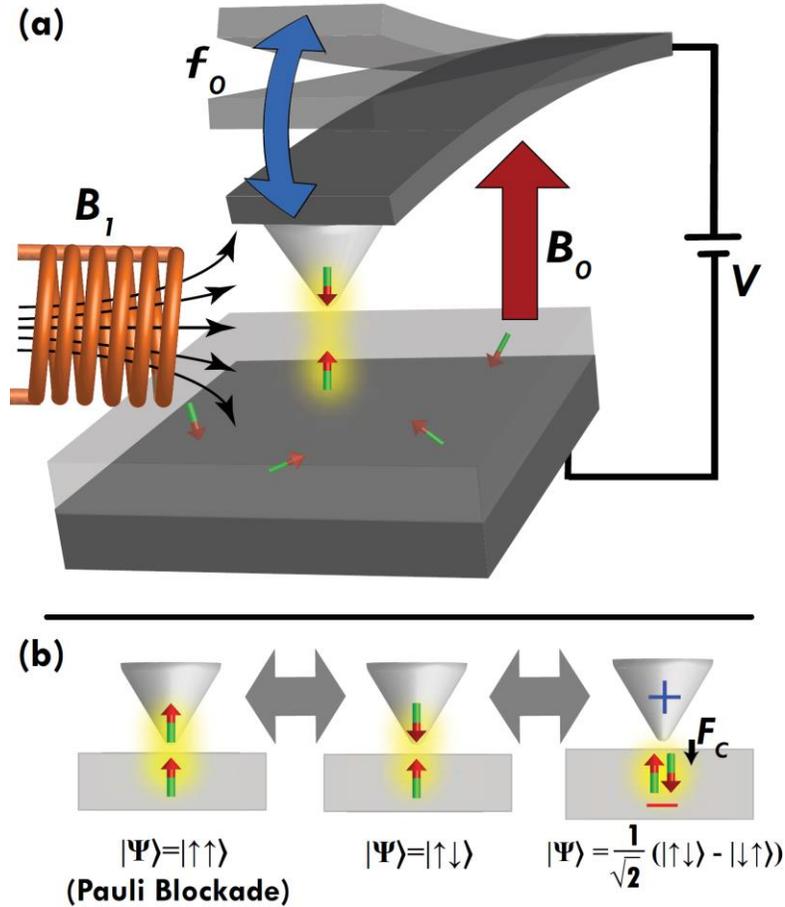

**Figure 1:** (a) Illustration of the proposed electrostatic force detected single spin microscope consisting of a scanning probe setup that includes a cantilever with a paramagnetic state at its tip, a paramagnetic state at the sample and a magnetic resonance setup (RF and DC magnetic fields). (b) Illustrations of three possible charge and spin configurations of the probe spin/test spin pair when energetic alignment and spatial proximity is achieved. Left: High triplet content when Pauli exclusion prohibits tunneling, but spin-relaxation allows for spin transitions towards mixed singlet/triplet states. Center: Spin pair states with mixed symmetry allow for tunneling. Right: Tunneling creates a doubly occupied diamagnetic singlet state where both the cantilever and the surface contain opposite charge whose net force gradient results in a cantilever frequency shift.

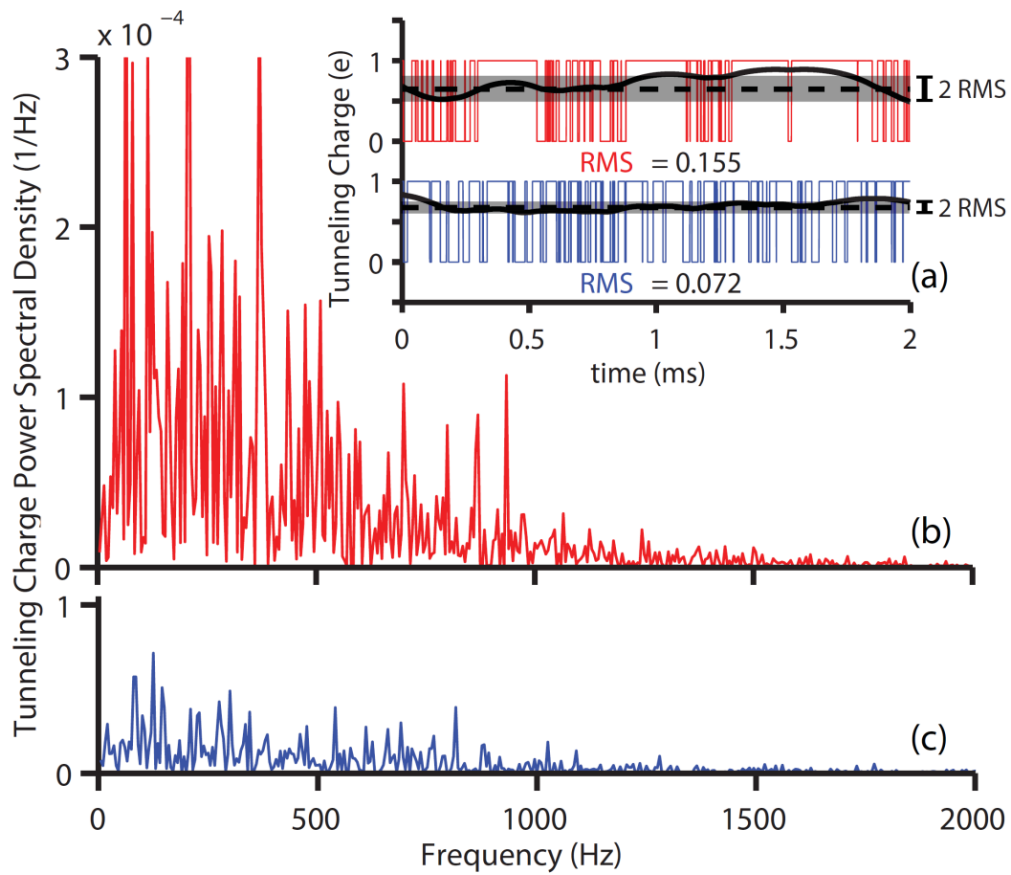

Figure 2: (a) Simulation of the charge power spectral density due to spin-dependent tunneling (RTS) between a paramagnetic cantilever state at the tip of a scanning probe cantilever (a probe spin) and a test spin at a sample surface in the absence (red) and presence (blue) of magnetic resonance conditions. The grey shaded areas around the average charge represent RMS values for the two cases obtained from the simulated data with an assumed detection bandwidth of 1kHz. For details on the simulation see supplemental material. (b,c) Plots of the tunneling charge power spectral density versus frequency obtained from the simulation when magnetic resonance is absent (b) and present (c). The spectral noise power density at lower frequency displays a significant reduction under magnetic resonance.

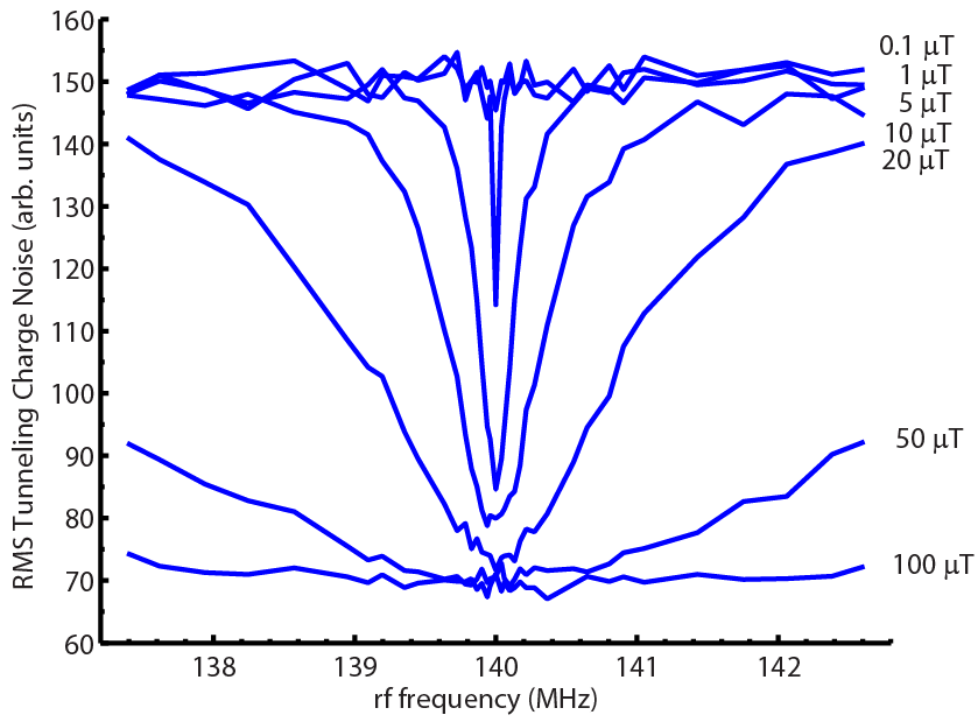

Figure 3: Simulation results of RMS tunneling charge noise as a function of rf frequency. Each resonance curve is produced with a 1s simulation time trace at each frequency step (40 frequency steps per resonance curve). The frequency steps are appropriately spaced in order to resolve the resonance. Error bars indicating the standard deviation of the simulation results (see Figs. 7 and 8) have been omitted for clarity in this plot. The different curves were simulated for different magnitudes of $B_1$. Each curve was simulated with an assumed $T_1$ = 200μs of the paramagnetic centers, a tunneling time $T_t$ = 10μs, and a detection bandwidth of 1000Hz.

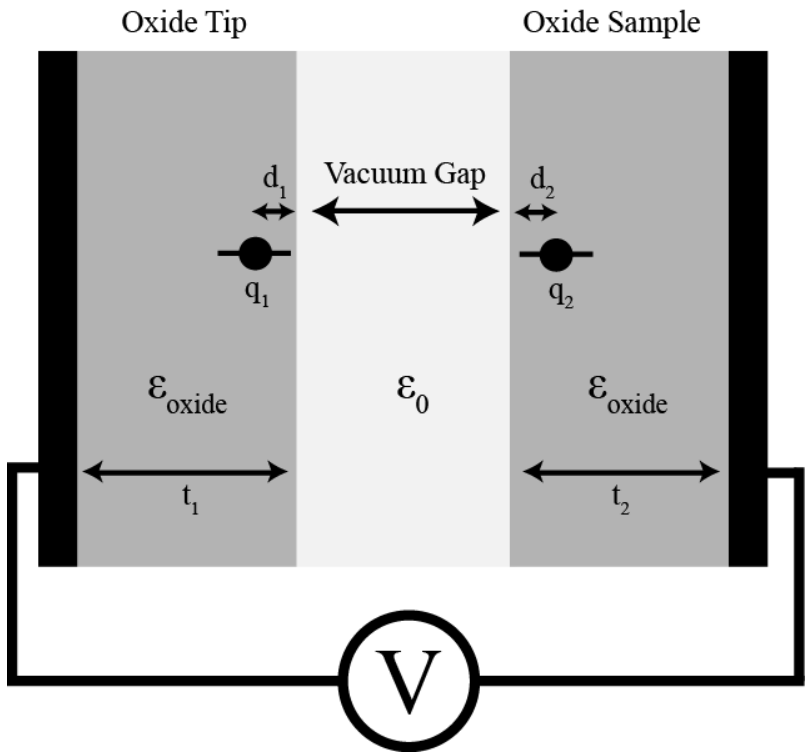

**Figure 4:** Electrostatic model showing a vacuum layer between two oxide layers with conducting back contacts. The electrostatic force gradient is calculated for two cases: 1) one electron in each defect state (separated charges), 2) electrons together in the doubly occupied defect state in the sample (charges together).

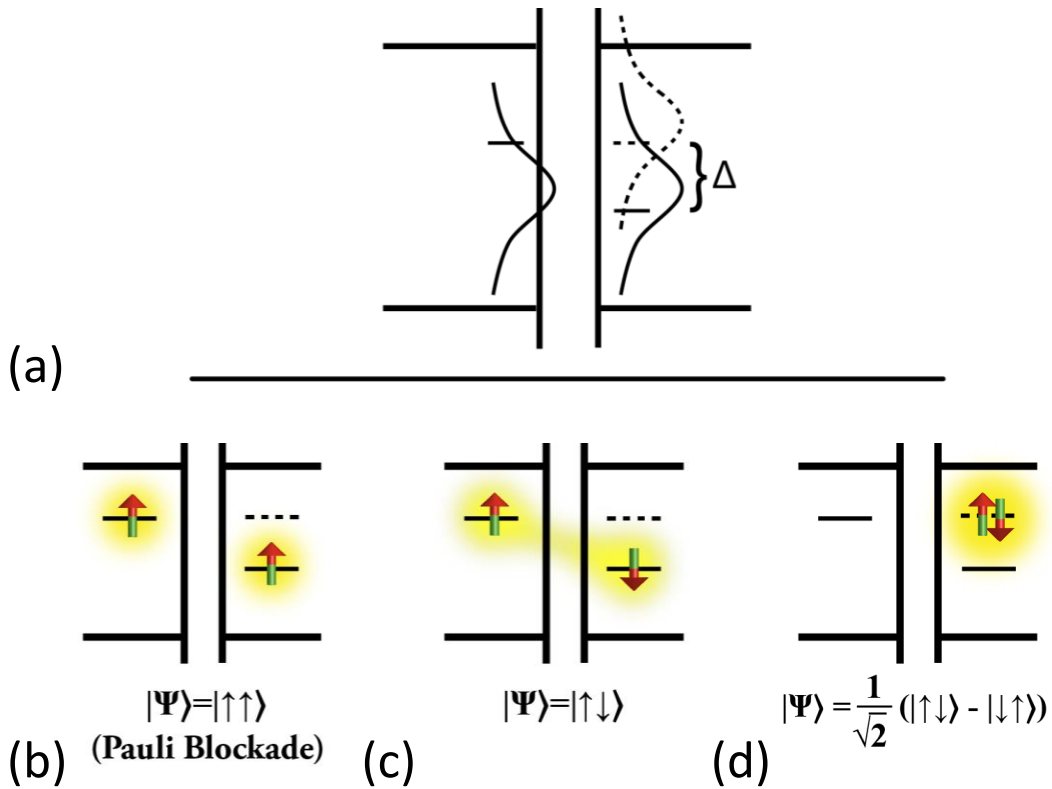

Figure 5: (a) Energy diagram of two paramagnetic defect states satisfying the energy condition for elastic tunneling. The tip state is on the left and the sample state is on the right. The horizontal solid/dashed lines represent the energetic location of the singly/doubly occupied states. The solid/dashed line Gaussian curves represent the energy spread of the singly/doubly occupied states. (b-d) Energy diagrams of the charge and spin configurations of the probe spin/test spin pair when energetic alignment and spatial proximity is achieved: (b) when tunneling is spin blocked, (c) when tunneling is possible, and (d) when the electrons doubly occupy one state in singlet configuration.

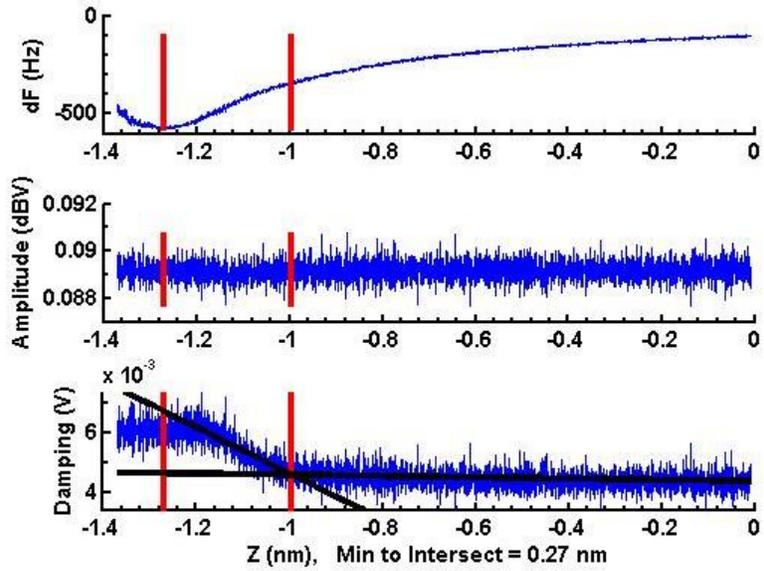

Figure 6: Plot of the measured AFM frequency shift, amplitude and damping signals versus tip-sample gap. The increase in the dissipation signal at z = 1 nm is attributed to the apex of the probe tip making first contact with the surface of the sample.

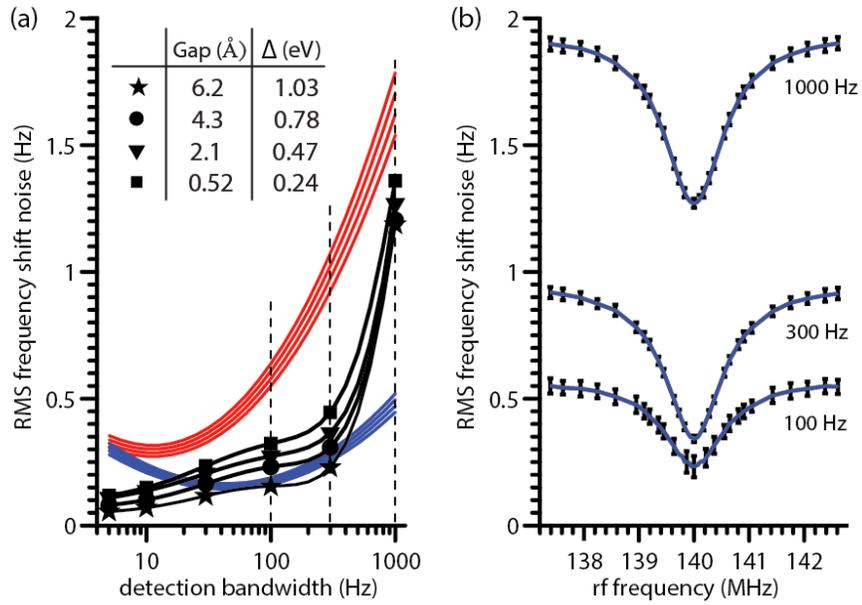

**Figure 7:** (a) Simulated frequency shift noise (RMS) caused by tunneling induced random telegraph noise in presence (blue) and absence (red) of magnetic resonance and measured system frequency shift noise (black symbols) connected by a guide to the eye (black line). All data where obtained for four different tip-sample gaps. The table shows the energy shift $\Delta$ (eV) of the probe and test state produced by an applied voltage of 10 volts at different tip-sample gaps. (b) Plot of the total frequency shift noise (RMS) consisting of simulated telegraph noise signal and the experimentally measured system noise levels as functions of the applied rf frequency for three bandwidth regimes at a tip-sample gap of 0.62 nm. For the assumed constant magnetic field of 5mT, the rf frequency range covers the *g*=2 electron spin resonance condition. The error bars indicate the standard deviation of the fluctuation of the simulated rts and measured noise power for an integration time of 1000ms. In order to discriminate on- from off-magnetic resonance conditions needed for the single spin detection, the on-resonance charge noise and the system noise need to be significantly lower than the off-resonance charge noise. This condition is fulfilled between ≈10 Hz and ≈1kHz bandwidth.

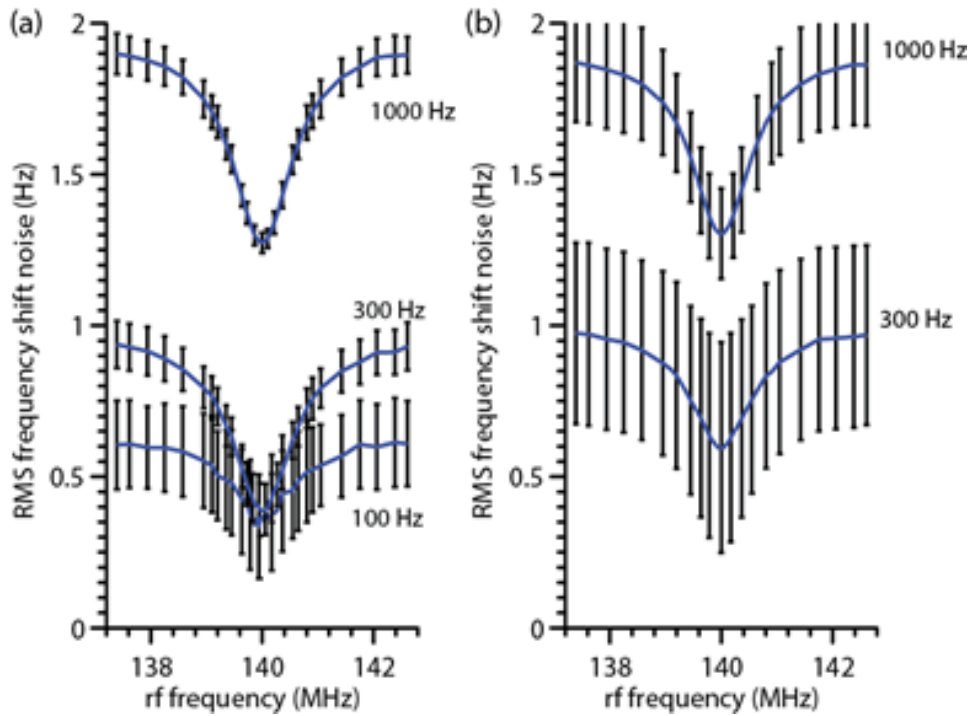

Figure 8: (a) Total RMS frequency shift noise, including both simulation and AFM system noise for 3 different bandwidths (1000 Hz, 300 Hz, 100 Hz) and as a function of rf frequency. This data is produced by simulations with a run time of 100ms per point. The error bars represent the predicted standard deviation of the measured noise due to the variance of the simulation noise and the calculated variance of the measured AFM system noise. (b) Same as in (a) but with a simulation time of 10 ms per frequency step with two bandwidths (1000 Hz and 300 Hz).